\documentclass[%
reprint,
 amsmath,amssymb,
 aps,
]{revtex4-2}

\usepackage{graphicx}
\usepackage{dcolumn}
\usepackage{bm}
\usepackage{multirow}
\usepackage{caption}
\usepackage{blindtext}

\usepackage[dvipsnames]{xcolor}
\usepackage{graphicx}
\begin{document}
\title{Exploring a novel model for projectile charge state distribution inside a solid-target}
\author{Soumya Chatterjee$^1$, Prashant Sharma$^2$,  D. Mitra$^1$ and T. Nandi$^{3*}$}
\affiliation{$^1$Department of Physics, University of Kalyani, Kalyani, West Bengal-741235, India.}
\affiliation{$^2$Department of Particle and Astrophysics, Weizmann Institute of Science, Rehovot 76100, Israel}
\affiliation{$^{3}$Inter-University Accelerator Centre, Aruna Asaf Ali Marg, Near Vasant Kunj, New Delhi-110067, India.}
\thanks {Email:\hspace{0.0cm} nanditapan@gmail.com. Present address: 1003 Regal, Mapsko Royal Ville, Sector-82, Gurgaon-122004, India.}
\begin{abstract}
For the first time, we report a theoretical methodology to predict charge state distribution of projectile ions inside a solid-target. The method utilizes either a simple Fermi gas model or an ab initio theoretical method and a certain parameterization of width for the Lorentzian charge state distributions. Results obtained from the two approaches are comparable, but the former has a certain edge over the latter. The projectile charge state distribution inside a solid-target plays a significant role in estimating electron capture cross-sections and then to describe the observed K-shell ionization dynamics. The electron capture process plays a certain role in L-shell ionization dynamics too, but in a test case of Si on Au target the subshell charge sharing contributes a more vital role than the electron capture. Thus, we have validated the present model as a reliable as well as useful for many solid-target based applications viz. tumour therapy, biophysics, accelerators, material science etc.
\end{abstract}

\maketitle

\section{Introduction}
\indent Though a monoenergetic beam with a fixed charge state is passed through the target material, a charge state distribution (CSD) of the projectile ions is manifested inside the target and being altered at the exit surface of the target as well as in flight because of autoionization. Let us call the former CSD occurring inside the target as CSD-I and the latter being altered at exit surface as well as in the flight outside the target as CSD-O. This CSD-O is measured with an electromagnetic device kept away from the target and is useful in accelerator technology, but the phenomena occurring inside or even at the exit surface require the CSD-I. However, this fact was not known and mistakenly, CSD-O had been used in place of CSD-I \cite{horvat2009ercs08} in studying inner shell ionization. Note that the innershell ionization of target atoms taking place inside the target, circular Rydberg state forming near the exit surface \cite{nandi2008formation}, ion energy-loss mostly happening inside the target and further modified in the exit surface \cite{nandi2013fast} etc. demands the knowledge of CSD-I.  Recently, a considerable difference between the CSD-I and CSD-O has been reported \cite{sharma2016experimental}. \\
\indent Inner shell ionization by ion impact has been investigated in the laboratory with the availability of the accelerators since the 1950's \cite{stier1954charge}. A vital role inside a target was put through a model associating the Auger processes which occur after the ions leave the solid \cite{betz1970charge}. Nevertheless, any direct measurement of the ionization phenomena inside the target was not possible until a couple of years ago \cite{nolen2013charge} using the x-ray spectroscopy technique \cite{sharma2016x}. The charge state of the projectile ion ($q$) in the beam-foil plasma created due to ion-solid interactions \cite{sharma2016experimental} is considerably higher than the measured ionic state outside the target \cite{sharma2019disentangling} because of electron capture phenomena from the exit surface \cite{nandi2008formation}. Such an 
interesting feature has not yet been exploited in the foil stripper technology till date. \\
\indent Inner shell ionization of target atoms are being carried out for quite a long time with light as well as heavy projectiles, for examples
\cite{benka1978tables,orlic1994experimental,kadhane2003k,lapicki2005status,zhou2013k,msimanga2016k,kumar2017shell,oswal2018x,hazim2020high,oswal2020experimental,miranda2020total}. 
It has enabled us to study the processes like ionization, excitation, multiple ionization \cite{lapicki2004effects}, radiative decay, Auger-decay \cite{dahl1976auger}, changes in atomic parameters, intrashell coupling effect \cite{pajek2003multiple,sarkadi1981possible}, etc at different energy regimes. Such processes occurring inside \cite{sharma2016x} as well as at the target surface \cite{nandi2008formation} change the initial charge state of the projectile to several charge states as measured by any set up placed away from the target. It is worth noting that the CSD depends on the initial parameters of the projectile ion (energy, initial charge state and atomic number) as well as target characteristics (thickness, density, and atomic number). Various groups have reported the CSDs using the techniques like the electromagnetic method \cite{MAIDIKOV1982295}, recoil separator \cite{LEINO1995653,KHUYAGBAATAR201240}, TOF \cite{DICKEL2015137} and CRBS \cite{SAADEH20112111} to obtain the CSDs of the projectile outside the target, which has the combined effect of charge exchange processes in the bulk as well as the surface of the target. However, these techniques fail to separately measure the CSD-I and CSD-O \cite{lifschitz2004effective}. Theoretical studies include only the CSD-O as seen in several reviews \cite{RevModPhys.30.1137, WITTKOWER1973113,SHIMA1986357,SHIMA1992173}, which include empirical models such as Bohr model \cite{bohr1941velocity}, Betz model \cite{betz1970charge}, Nikolaev-Dmitriev model \cite{nikolaev1968equilibrium}, To-Drouin model \cite{to1976etude}, Shima-Ishihara-Mikumo model \cite{shima1982empirical}, Itoh model \cite{itoh1999equilibrium}, 
and Schiwietz model \cite{schiwietz2004femtosecond}. However, no effort is yet put on developing any theoretical model for estimating the CSD-I because of no experimental triumph till date. \\
\begin{figure*}
\centering
\includegraphics[width=18cm,height=6cm]{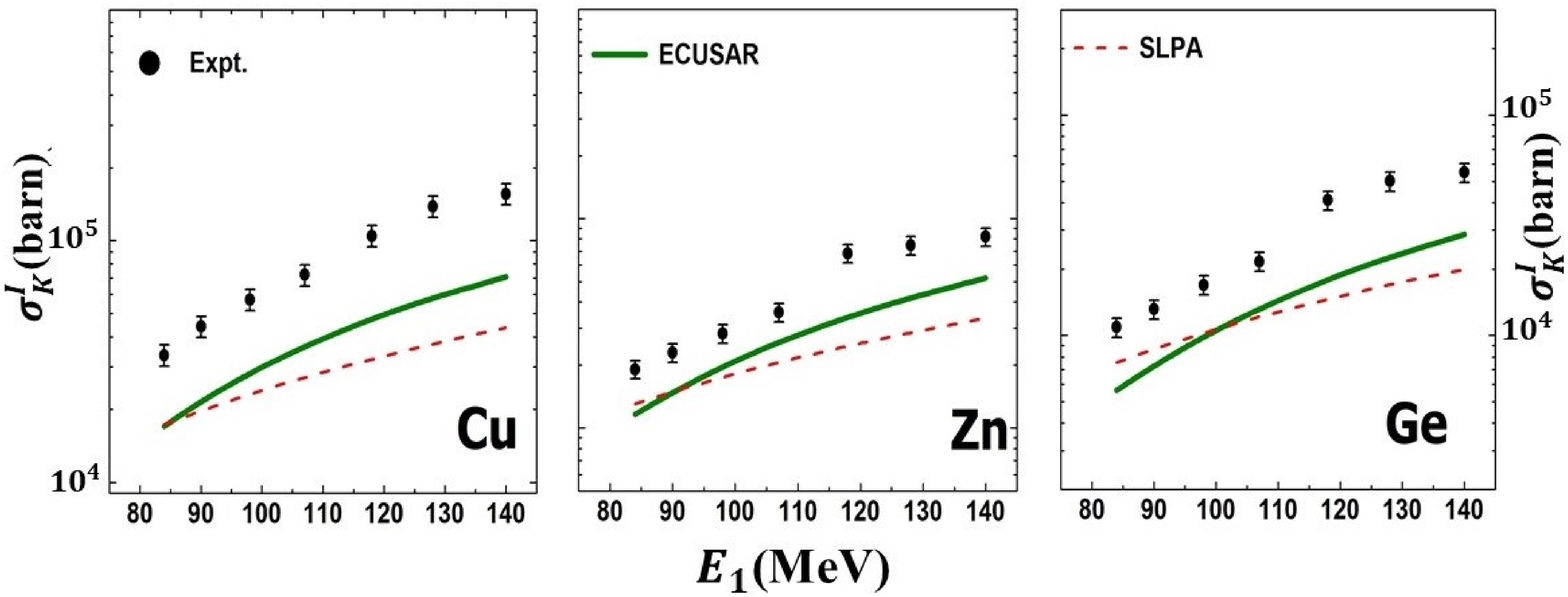}
\caption{Comparison of experimental K shell ionization cross sections of different targets bombarded by the $^{28}{Si}$ ions as a function of ion-beam energies with the direct ionization cross sections from ECUSAR \cite{lapicki2004effects} and SLPA \cite{montanari2013theory}. Note that simultaneous multiple ionization effects were well accounted during evaluation of K-shell ionization cross-sections from measured K-x ray production cross-sections \cite{chatterjee2021investigation}.
\label{Expt and DI theories}}
\end{figure*}
\indent In the present letter, we have demonstrated that the K-shell as well as L-subshell ionization cross-sections of the target atoms by heavy-ion impact can be well described if the CSD-I using a novel Fermi gas model and  a certain parameterization of width for the Lorentzian charge state distribution is taken into account in calculating the electron capture cross-sections. The CSD-I finds significant applications in tumour therapy, biophysics, accelerators, material science etc.
\section{Background}
\indent Recently, measured the heavy induced K-shell ionization cross-section ($\sigma^I_{K}$) \cite{chatterjee2021investigation}, where simultaneous multiple ionization effects were accounted while evaluating the ionization cross-sections from production cross-sections and currently most accurate fluorescence data \cite{krause1979atomic} were used. In spite of these, we can notice from Fig.\ref{Expt and DI theories} that the measured $\sigma^I_{K}$ are at least a factor of two times higher than the predictions of the ECUSAR \cite{lapicki2004effects} and SLPA \cite{montanari2011collective} theories.\\
\indent The higher experimental ionization probability than the theoretical prediction provides a clear indication that the direct (Coulomb) ionization is not at all enough to describe the K-shell ionization phenomenon. In addition to direct ionization, another parallel mechanism must be involved. Such a possibility can arise from K-K electron capture phenomena. It can only be feasible if the K-shell of the projectile is either fully or partially vacant, i.e., for the Si-ion impact case, $Si^{13+}$ and $Si^{14+}$ must be present inside the target. If such condition is met, the target K electron may be captured by the vacant K shell of the projectile ions and then the K shell ionization of the target atoms will increase due to the K-K capture process along with the Coulomb ionization.\\
\section{Theory of electron capture phenomenon}
\indent To calculate the K-K electron capture cross sections, we can use the theory of \citet{lapicki1977electron} which is based on the Oppenheimer-Brinkman-Kramers (OBK) approximation \cite{may1964formation} with binding and Coulomb deflection corrections at low velocities. Neglecting the change in the binding energy of the K shell electron of the projectile with one versus two K shell vacancies, a statistical scaling is used to calculate the electron transfer cross section for the case of one projectile K-shell vacancy, $\sigma_{1K \rightarrow K}$ to be $\sigma_{2K \rightarrow  K}/2$, $\sigma_{2K \rightarrow K}$=production cross section for two projectile K-shell vacancy. In the present experimental condition, $v_1= 10.96-14.15$ and $v_{2k}= 28.7 - 31.7$ atomic units, thus $\sigma_{2K \rightarrow  K}$ can be chosen from Ref.\cite{lapicki1977electron} as follows
\begin{equation}
\sigma_{2k \rightarrow k}=\frac{1}{3}\sigma_{2k\rightarrow k}^{OBK}(\theta_k),\theta_k=\frac{E_K}{Z_{2k}^2\times13.6}\label{KKCap} 
\end{equation}
\noindent with $Z_{2k}=Z_2-0.3$ and
\begin{equation}
\sigma_{2k\rightarrow k}^{OBK}(\theta_k)= \frac{2^9}{5v_1^2}\pi a_0^2 \frac{(v_{1k}v_{2K})^5}{[v_{1k}^2+(v_1^2+v_{2k}^2-v_{1k}^2)^2/4v_1^2]^5}.  
\end{equation}
Here the subscripts 1 and 2 indicate the projectile ion and target atom, respectively,  $a_0$, $v_{1K}$, and $v_{2K}$ are Bohr radius, K-shell orbital velocity for the projectile ion and target atom, respectively. And $E_K$ is the binding energy of the K-shell electron of the target in eV. Note that this K-K capture theory has been adapted to estimate the K-L and L-K capture cross sections too.\\
\begin{figure*}
\centering
\includegraphics[width=17.5cm,height=16.0 cm]{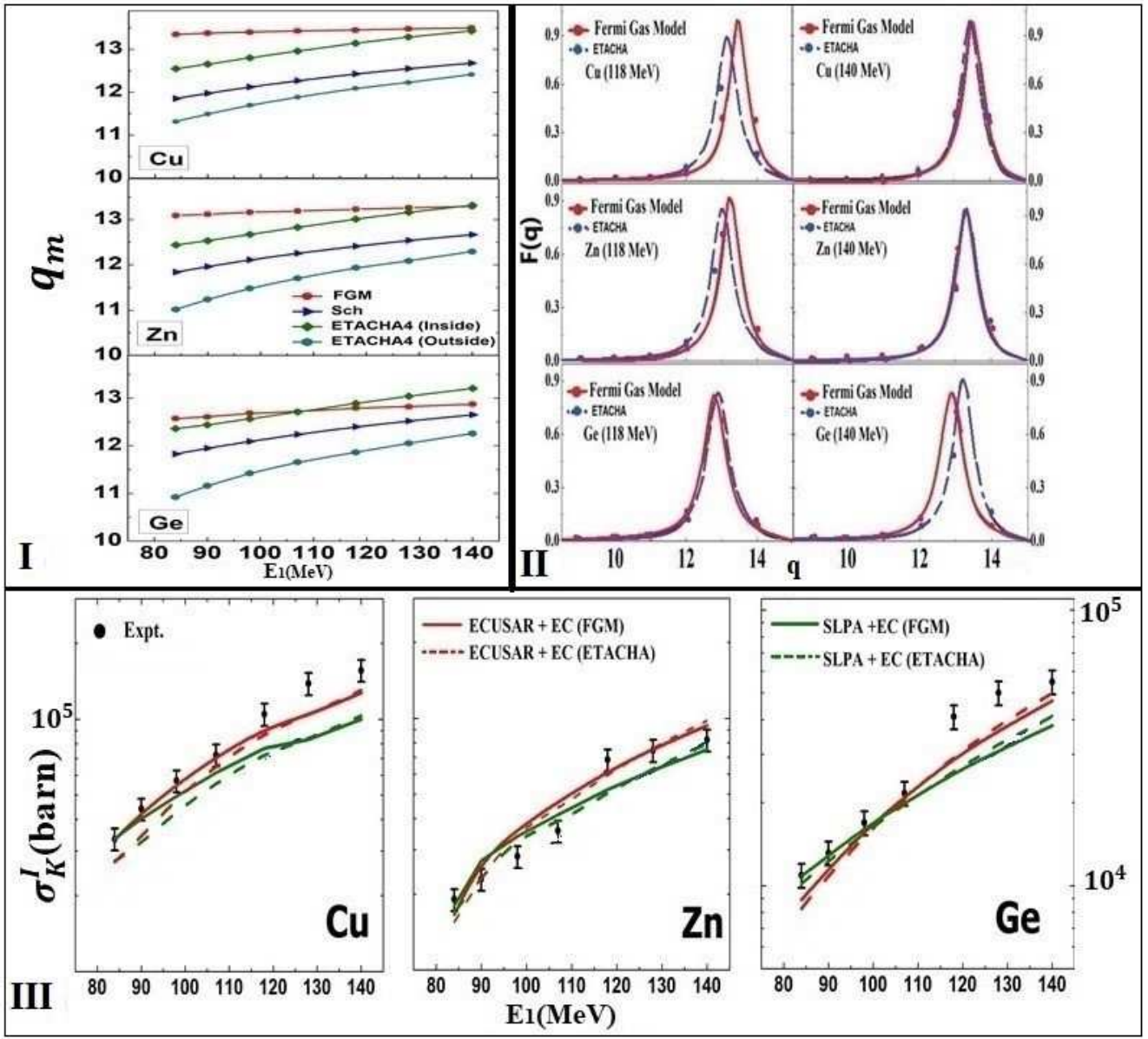}
\caption{I. Mean charge state of $^{28}{Si}$ ions inside the targets ${Cu}$, ${Zn}$ and ${Ge}$ as predicted by the Fermi Gas Model (FGM) \cite{brandt1973dynamic} and the same outside the targets as predicted by the Schiwietz model (Sch) \cite{schiwietz2001improved} versus the incident energies. II. Charge state distributions in side different targets according to the Fermi Gas Model (FGM) \cite{brandt1973dynamic} and ETACHA4 (inside) \cite{ETACHA4} at two select energies. III. Comparison of experimental K shell ionization cross sections with the ECUSAR (DI) \cite{lapicki2004effects} and the sum of the ECUSAR (DI) and K-K capture cross section (EC) \cite{lapicki1977electron} predictions for different targets bombarded by the $^{28}{Si}$ ions are given as a function of ion-beam energies.}
\label{Q_LORENTZ}
\end{figure*}
\indent Due to certain CSD-I, effective K-K capture contribution will be $F(q)\times \sigma_{2k\rightarrow k}^{OBK}(\theta_k)$, for $q$=13 and 14. Here $F(q)$ is the charge state fraction for a specific $q$. To obtain the $F(q)$, in the first step, we have used the following methods: (i) ab initio approach by means of ETACHA4 code \cite{ETACHA4} and (ii) Fermi gas model based empirical formula \cite{brandt1973dynamic}. Note that about a decade ago, significance of CSD-I on the target ionization was not at all known. Thus, ERCS08 code \citet{horvat2009ercs08} has made use of CSD-O, which is fully incorrect as be evident after a while. \\
\indent Recently, Lamour \textit{et. al.} \cite{ETACHA4} developed a computer code called ETACHA4 to compute the charge state fractions of the projectile ions on the passage of a target medium, either solid or gas, by employing suitable rate equations. In the code, the non-radiative and radiative electron capture cross-sections are calculated using the relativistic eikonal approximation \cite{Meyerhof} and Bethe-Salpeter formula \cite{bethe1957quantum}, respectively. The total electron capture cross-section is sum of the non-radiative and radiative electron capture cross-sections. Whereas, the ionization and excitation cross-sections are estimated using the continuum distorted-wave-eikonal initial state approximation \cite{fainstein1987z,fainstein1991two} and symmetric eikonal model \cite{olivera1993electronic,ramirez1995distortion}, respectively. Significant contribution of  the non-radiative and radiative electron capture remains intact that takes place at the exit layers because that forming inside the solid-target are destroyed in the following collisions if geometrical size  of the excited stated states so created is larger than the lattice parameter of the target material. This is the reason,  the excited-state formation occurs at the exit surface \cite{tolk1981role}.  Hence, putting the electron capture cross section equals to zero in the calculation using the ETACHA4 code provide us a good estimate of the CSD-I.\\ 
\indent According to the  Fermi gas model based empirical formula, the mean charge state ($q_m$) inside the target \cite{brandt1973dynamic} is given by
\begin{equation}
q_m = z_1(1- {v_F\over v_1})
\end{equation}
where $ z_1$ and $v_F$ are the projectile atomic number and Fermi velocity of target electrons, respectively. $v_F$ of $Cu, Zn$ and $Ge$ are $1.11\times10^6$, $1.566\times10^6$ \cite{gall2016electron} and $2.5\times10^6$ m/s \cite{isaacson1975compilation}, respectively. To showcase the difference of ionization of the projectile ion inside and outside the target, we have displayed the  $q_m $ as predicted by the Fermi gas model \cite{brandt1973dynamic} and Schiwietz model \cite{schiwietz2001improved} in Fig.\ref{Q_LORENTZ}(I). This contrasting picture is governed by the solid surface \cite{nandi2008formation,sharma2019disentangling}. \\
\indent In second step, the  $q_m$-values inside the target are substituted in the Lorentzian charge state distribution \cite{sharma2016experimental} to obtain the F(q) as follows
\begin{equation}
F(q)=\frac{1}{\pi}\frac{\frac{\Gamma}{2}}{(q-q_m)^2+(\frac{\Gamma}{2})^2} \:\text{and}\:\sum_q F(q)=1 \label{CSF}
\end{equation}
\noindent here distribution width $\Gamma$ is taken from Novikov and Teplova \cite{novikov2014method} as follows
\begin{equation}
\centering
\Gamma(x)= C[1-exp(-(x)^\alpha)][1-exp(-(1-x)^\beta)]\label{Gamma}
\end{equation}
\noindent where $x=q_m/z,\:\alpha=0.23,\: \beta=0.32 \hspace{0.2cm}$ and $C=2.669-0.0098× Z_2+0.058×Z_1+0.00048×Z_1× Z_2$. The $F(q)$ values so obtained are shown in Fig.\ref{Q_LORENTZ}(II). Further, Charge state fraction [F(q)] of Si$^{13+}$ and Si$^{14+}$ obtained from FGM \cite{brandt1973dynamic} and ETACHA4 \cite{ETACHA4} in different target elements and at various kinetic energies of Si-ion beam. Note that FGM and ETACHA4 represent F(q) inside the target.\\ 
\begin{figure}
\centering
\includegraphics[width=8.5cm,height=18.5cm]{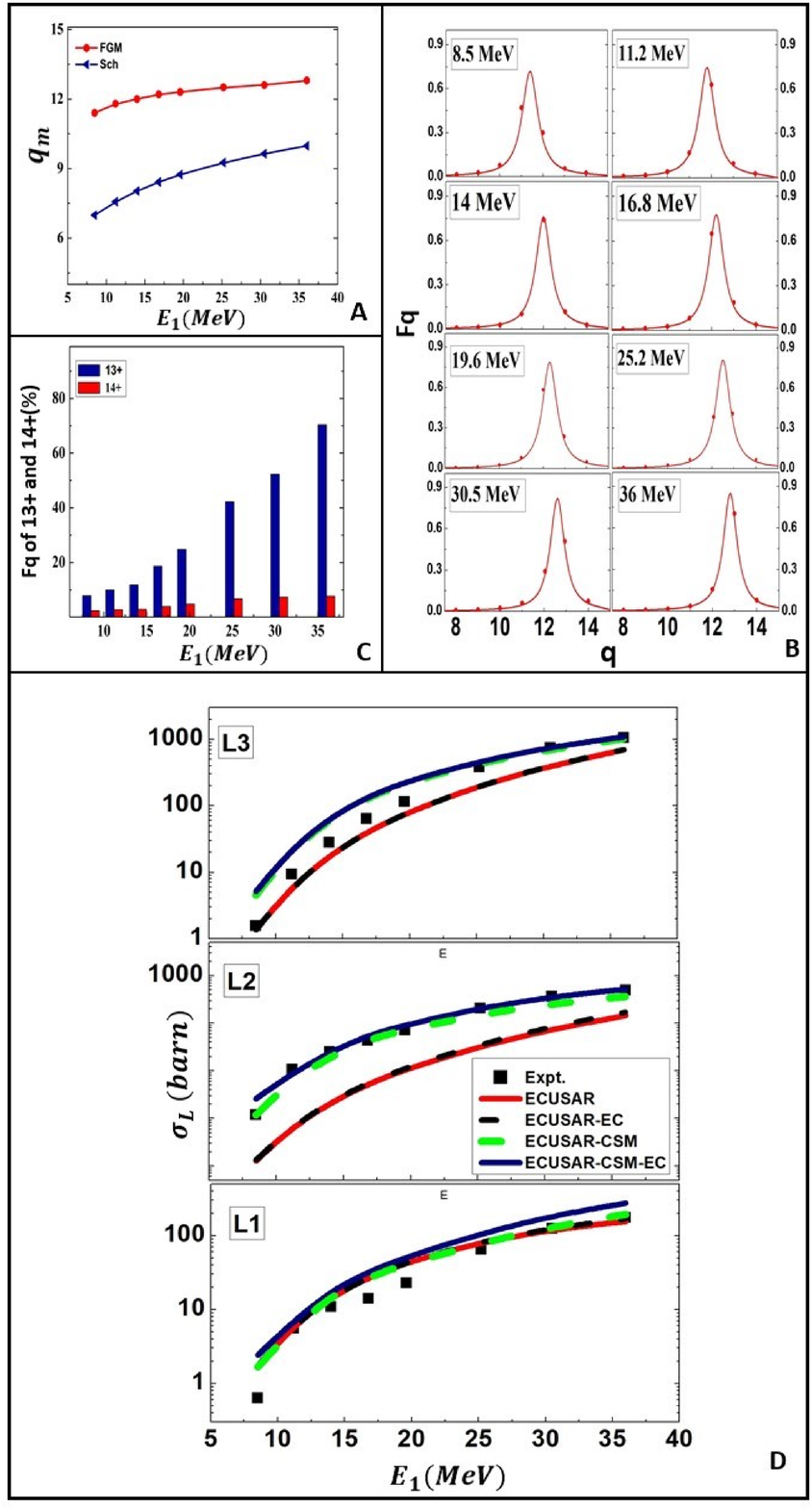}
\caption{A. Mean charge state of $^{28}{Si}$ ions inside the target ${Au}$ as predicted by the Fermi Gas Model (FGM) \cite{brandt1973dynamic} and the same outside the target as obtained from the Schiwietz model (Sch) \cite{schiwietz2001improved} versus the incident energies. B. Charge state distributions inside the target at different energies, C. Charge state fraction (F(q)) chart for q= 13+ and 14+ inside the target as a function of the beam energies using the FGM, and D. Comparison of the measured L-subshell ionization cross-sections with different theories due to bombardment of $^{28}Si$ ions on Au target at different incident energies.}
\label{Si_Au}
\end{figure}
\section{Results and discussions}
\indent Mean charge state inside and outside the target as a function of beam energy is depicted in Fig.\ref{Q_LORENTZ}(I). \citet{schiwietz2001improved} formula is used to obtain the mean charge state outside the target, whereas mean charge state inside the target is taken either form \cite{ETACHA4} or \cite{brandt1973dynamic}.
Here we can notice the predictions for mean charge state inside the target from ETACHA4 and FGM are quite close, but a large difference between the mean charge state inside and outside the target and thus considering CSD-O for innershell ionization studies is not at all justified. CSD-I from ETACHA4 and FGM are also quite close as shown in Fig.\ref{Q_LORENTZ}(II). This charge state fractions are utilized in calculation of electron capture cross-section. Now, the $\sigma_{KK}$  and $\sigma_{KL}$ so obtained are added with the $\sigma^I_{K}$ as obtained from direct ionization theories and plotted in Fig.\ref{Q_LORENTZ}(III). At this instance, excitation function curve of the sum of direct and K-K + K-L capture cross sections show a good agreement with the corresponding experimental cross sections. Furthermore, ECUSAR - FGM gives the closer agreement with the experimental data. Such agreement reveals that a simple Fermi gas model gives a correct estimation of the $q_m$ inside the target, where $v_F$ plays a central role and needs accurate evaluation.\\ 
\indent The direct ionization cross section alone does not describe the K-shell ionization phenomenon in the present collision. Electron capture processes are equally important and electron capture cross section is calculated using the charge state fractions from different theoretical methods. Here, we see that the electron capture cross sections are almost equal \cite{chatterjee2021investigation} if the charge state fractions are taken from FGM \cite{brandt1973dynamic} and ETACHA4 \cite{ETACHA4}. Therefore, either FGM or ETACHA4 can be used to estimate the projectile charge state distribution inside the solid-target. However, ETACHA4 can handle up to a certain number of electrons in the projectile ion and thus having difficulties for heavy projectiles. Whereas no such restrictions are there with the FGM and thus we shall use only the FGM for next application.\\ 
\indent This approach so developed has been validated through earlier experimental results of Si ion on Au \cite{fijal2008coupling} and shown in Fig. \ref{Si_Au}. Fermi velocity of Au is taken as 1.40$\times 10^6$ m/s \cite{isaacson1975compilation}. Fig. \ref{Si_Au} A shows the large difference between the mean charge state of the projectile ions inside and outside the target in the lower energy range than that shown in Fig. \ref{Q_LORENTZ} (I). The CSD-I is depicted at different energies in Fig. \ref{Si_Au} B. Corresponding charge state factions for $q=13+$ and $14+$ are shown in Fig. \ref{Si_Au} C, which are used in the calculation of the L-K capture cross sections \cite{lapicki1977electron}. Such L-K capture cross sections are added to direct ionization cross-section using ECUSAR as well as ECUSAR-CSM. CSM stands for the coupled subshell mechanism, which evolves from the concept of the charge sharing among the L-subshells \cite{fijal2008coupling}.
Finally, different theoretical values of L-subshell ionization cross-sections are compared with the measured values in Fig. \ref{Si_Au} D. This figure shows that the ECUSAR-EC does not make much effect, where as the ECUSAR-CSM is quite close to the measured cross-sections. In contrast, the electron capture process plays a significant role in K-shall ionization as shown in Fig.\ref{Q_LORENTZ}(III). Therefore, the electron capture is an important phenomenon in heavy-ion impact ionization processes. Even though its contribution is small in Si-induced Au target ionization, but this contribution may be profound for higher $Z_1/Z_2$-values as shown in above mentioned K-shell ionization cases, $Z_1$ and $Z_2$ are the atomic number  of the projectile ions and target atoms, respectively. Hence, in L-subshell ionization both the charge sharing as well as electron capture processes are coexistent and may be competing with each other, especially, when $Z_1$ tends to $Z_2$. Such measurements are very rare. Our literature survey did not lead to find any L-subshell ionization cross-section measurements as a function of beam energy due to bombardment of heavy ions with $Z>14$.\\

\section{Conclusion}
\indent We have developed a novel formula based on Fermi gas model that predicts very well the mean charge state of the projectile ions inside the solid-target. Further, we present a methodology that generates the charge state distribution of the projectile ions inside the solid-target using the mean charge state so obtained, Lorentzian distribution and its width from a standard formula. Such charge state distribution is utilized in calculating the electron capture cross-section and sum of this capture cross-section and direct ionization cross-section describe the measured ionization cross-sections for K-shell very well. However, the sum of the electron capture and direct ionization does not describe the L-subshell ionization phenomenon, we  require the charge sharing among the subshells too. For the collision system Si-ion on Au-target the charge sharing process is more vital than the electron capture. This is certainly an important topic of research needing further attention from both the experimental as well as theoretical studies.  On the application side, the mean charge state  as well as  charge state distribution of the heavy projectile ions inside the solid-target find important applications in tumour therapy, biophysics, accelerators, material science etc. \\
%
\indent We have shown that the mean charge state inside the foil is much higher than that outside it because the electron capture process at the exit surface plays an significant role in reducing the charge state from a higher value to a lower one for the low to medium energy ions. We may obtain a high charge state from a foil stripper if the electron capture phenomena at the stripper foil surface is restricted by a special surface engineering, by converting the surface from conducting to insulating one. Once such an idea is realized a renaissance will take place in radioactive ion beam research \cite{grigorenko2019scientific} and  accelerator technology \cite{adonin2014beam}. Hence, the simple Fermi gas model imparts a significant contribution to the heavy-ion collisions with the solid-targets.\\
\indent One of the authors, SC, acknowledges the University of Kalyani for providing him the senior research fellowship towards obtaining PhD.
\\
\bibliography{main.bbl}

\begin{thebibliography}{62}%
\makeatletter
\providecommand \@ifxundefined [1]{%
 \@ifx{#1\undefined}
}%
\providecommand \@ifnum [1]{%
 \ifnum #1\expandafter \@firstoftwo
 \else \expandafter \@secondoftwo
 \fi
}%
\providecommand \@ifx [1]{%
 \ifx #1\expandafter \@firstoftwo
 \else \expandafter \@secondoftwo
 \fi
}%
\providecommand \natexlab [1]{#1}%
\providecommand \enquote  [1]{``#1''}%
\providecommand \bibnamefont  [1]{#1}%
\providecommand \bibfnamefont [1]{#1}%
\providecommand \citenamefont [1]{#1}%
\providecommand \href@noop [0]{\@secondoftwo}%
\providecommand \href [0]{\begingroup \@sanitize@url \@href}%
\providecommand \@href[1]{\@@startlink{#1}\@@href}%
\providecommand \@@href[1]{\endgroup#1\@@endlink}%
\providecommand \@sanitize@url [0]{\catcode `\\12\catcode `\$12\catcode
  `\&12\catcode `\#12\catcode `\^12\catcode `\_12\catcode `\%12\relax}%
\providecommand \@@startlink[1]{}%
\providecommand \@@endlink[0]{}%
\providecommand \url  [0]{\begingroup\@sanitize@url \@url }%
\providecommand \@url [1]{\endgroup\@href {#1}{\urlprefix }}%
\providecommand \urlprefix  [0]{URL }%
\providecommand \Eprint [0]{\href }%
\providecommand \doibase [0]{http://dx.doi.org/}%
\providecommand \selectlanguage [0]{\@gobble}%
\providecommand \bibinfo  [0]{\@secondoftwo}%
\providecommand \bibfield  [0]{\@secondoftwo}%
\providecommand \translation [1]{[#1]}%
\providecommand \BibitemOpen [0]{}%
\providecommand \bibitemStop [0]{}%
\providecommand \bibitemNoStop [0]{.\EOS\space}%
\providecommand \EOS [0]{\spacefactor3000\relax}%
\providecommand \BibitemShut  [1]{\csname bibitem#1\endcsname}%
\let\auto@bib@innerbib\@empty
\bibitem [{\citenamefont {Horvat}(2009)}]{horvat2009ercs08}%
  \BibitemOpen
  \bibfield  {author} {\bibinfo {author} {\bibfnamefont {V.}~\bibnamefont
  {Horvat}},\ }\href@noop {} {\bibfield  {journal} {\bibinfo  {journal}
  {Computer Physics Communications}\ }\textbf {\bibinfo {volume} {180}},\
  \bibinfo {pages} {995} (\bibinfo {year} {2009})}\BibitemShut {NoStop}%
\bibitem [{\citenamefont {Nandi}(2008)}]{nandi2008formation}%
  \BibitemOpen
  \bibfield  {author} {\bibinfo {author} {\bibfnamefont {T.}~\bibnamefont
  {Nandi}},\ }\href@noop {} {\bibfield  {journal} {\bibinfo  {journal} {The
  Astrophysical Journal Letters}\ }\textbf {\bibinfo {volume} {673}},\ \bibinfo
  {pages} {L103} (\bibinfo {year} {2008})}\BibitemShut {NoStop}%
\bibitem [{\citenamefont {Nandi}\ \emph {et~al.}(2013)\citenamefont {Nandi},
  \citenamefont {Haris}, \citenamefont {Singh}, \citenamefont {Kumar},
  \citenamefont {Kumar}, \citenamefont {Saini}, \citenamefont {Khan},
  \citenamefont {Jhingan}, \citenamefont {Verma}, \citenamefont {Tauheed} \emph
  {et~al.}}]{nandi2013fast}%
  \BibitemOpen
  \bibfield  {author} {\bibinfo {author} {\bibfnamefont {T.}~\bibnamefont
  {Nandi}}, \bibinfo {author} {\bibfnamefont {K.}~\bibnamefont {Haris}},
  \bibinfo {author} {\bibfnamefont {G.}~\bibnamefont {Singh}}, \bibinfo
  {author} {\bibfnamefont {P.}~\bibnamefont {Kumar}}, \bibinfo {author}
  {\bibfnamefont {R.}~\bibnamefont {Kumar}}, \bibinfo {author} {\bibfnamefont
  {S.}~\bibnamefont {Saini}}, \bibinfo {author} {\bibfnamefont
  {S.}~\bibnamefont {Khan}}, \bibinfo {author} {\bibfnamefont {A.}~\bibnamefont
  {Jhingan}}, \bibinfo {author} {\bibfnamefont {P.}~\bibnamefont {Verma}},
  \bibinfo {author} {\bibfnamefont {A.}~\bibnamefont {Tauheed}},  \emph
  {et~al.},\ }\href@noop {} {\bibfield  {journal} {\bibinfo  {journal}
  {Physical review letters}\ }\textbf {\bibinfo {volume} {110}},\ \bibinfo
  {pages} {163203} (\bibinfo {year} {2013})}\BibitemShut {NoStop}%
\bibitem [{\citenamefont {Sharma}\ and\ \citenamefont
  {Nandi}(2016{\natexlab{a}})}]{sharma2016experimental}%
  \BibitemOpen
  \bibfield  {author} {\bibinfo {author} {\bibfnamefont {P.}~\bibnamefont
  {Sharma}}\ and\ \bibinfo {author} {\bibfnamefont {T.}~\bibnamefont {Nandi}},\
  }\href@noop {} {\bibfield  {journal} {\bibinfo  {journal} {Physics of
  Plasmas}\ }\textbf {\bibinfo {volume} {23}},\ \bibinfo {pages} {083102}
  (\bibinfo {year} {2016}{\natexlab{a}})}\BibitemShut {NoStop}%
\bibitem [{\citenamefont {Stier}\ \emph {et~al.}(1954)\citenamefont {Stier},
  \citenamefont {Barnett},\ and\ \citenamefont {Evans}}]{stier1954charge}%
  \BibitemOpen
  \bibfield  {author} {\bibinfo {author} {\bibfnamefont {P.}~\bibnamefont
  {Stier}}, \bibinfo {author} {\bibfnamefont {C.}~\bibnamefont {Barnett}}, \
  and\ \bibinfo {author} {\bibfnamefont {G.}~\bibnamefont {Evans}},\
  }\href@noop {} {\bibfield  {journal} {\bibinfo  {journal} {Physical Review}\
  }\textbf {\bibinfo {volume} {96}},\ \bibinfo {pages} {973} (\bibinfo {year}
  {1954})}\BibitemShut {NoStop}%
\bibitem [{\citenamefont {Betz}\ and\ \citenamefont
  {Grodzins}(1970)}]{betz1970charge}%
  \BibitemOpen
  \bibfield  {author} {\bibinfo {author} {\bibfnamefont {H.~D.}\ \bibnamefont
  {Betz}}\ and\ \bibinfo {author} {\bibfnamefont {L.}~\bibnamefont
  {Grodzins}},\ }\href@noop {} {\bibfield  {journal} {\bibinfo  {journal}
  {Physical Review Letters}\ }\textbf {\bibinfo {volume} {25}},\ \bibinfo
  {pages} {211} (\bibinfo {year} {1970})}\BibitemShut {NoStop}%
\bibitem [{\citenamefont {Nolen}\ and\ \citenamefont
  {Marti}(2013)}]{nolen2013charge}%
  \BibitemOpen
  \bibfield  {author} {\bibinfo {author} {\bibfnamefont {J.~A.}\ \bibnamefont
  {Nolen}}\ and\ \bibinfo {author} {\bibfnamefont {F.}~\bibnamefont {Marti}},\
  }\href@noop {} {\bibfield  {journal} {\bibinfo  {journal} {Reviews of
  Accelerator Science and Technology}\ }\textbf {\bibinfo {volume} {6}},\
  \bibinfo {pages} {221} (\bibinfo {year} {2013})}\BibitemShut {NoStop}%
\bibitem [{\citenamefont {Sharma}\ and\ \citenamefont
  {Nandi}(2016{\natexlab{b}})}]{sharma2016x}%
  \BibitemOpen
  \bibfield  {author} {\bibinfo {author} {\bibfnamefont {P.}~\bibnamefont
  {Sharma}}\ and\ \bibinfo {author} {\bibfnamefont {T.}~\bibnamefont {Nandi}},\
  }\href@noop {} {\bibfield  {journal} {\bibinfo  {journal} {Physics Letters
  A}\ }\textbf {\bibinfo {volume} {380}},\ \bibinfo {pages} {182} (\bibinfo
  {year} {2016}{\natexlab{b}})}\BibitemShut {NoStop}%
\bibitem [{\citenamefont {Sharma}\ and\ \citenamefont
  {Nandi}(2019)}]{sharma2019disentangling}%
  \BibitemOpen
  \bibfield  {author} {\bibinfo {author} {\bibfnamefont {P.}~\bibnamefont
  {Sharma}}\ and\ \bibinfo {author} {\bibfnamefont {T.}~\bibnamefont {Nandi}},\
  }\href@noop {} {\bibfield  {journal} {\bibinfo  {journal} {Physical Review
  Accelerators and Beams}\ }\textbf {\bibinfo {volume} {22}},\ \bibinfo {pages}
  {034501} (\bibinfo {year} {2019})}\BibitemShut {NoStop}%
\bibitem [{\citenamefont {Benka}\ and\ \citenamefont
  {Kropf}(1978)}]{benka1978tables}%
  \BibitemOpen
  \bibfield  {author} {\bibinfo {author} {\bibfnamefont {O.}~\bibnamefont
  {Benka}}\ and\ \bibinfo {author} {\bibfnamefont {A.}~\bibnamefont {Kropf}},\
  }\href@noop {} {\bibfield  {journal} {\bibinfo  {journal} {Atomic Data and
  Nuclear Data Tables}\ }\textbf {\bibinfo {volume} {22}},\ \bibinfo {pages}
  {219} (\bibinfo {year} {1978})}\BibitemShut {NoStop}%
\bibitem [{\citenamefont {Orlic}\ \emph {et~al.}(1994)\citenamefont {Orlic},
  \citenamefont {Sow},\ and\ \citenamefont {Tang}}]{orlic1994experimental}%
  \BibitemOpen
  \bibfield  {author} {\bibinfo {author} {\bibfnamefont {I.}~\bibnamefont
  {Orlic}}, \bibinfo {author} {\bibfnamefont {C.}~\bibnamefont {Sow}}, \ and\
  \bibinfo {author} {\bibfnamefont {S.}~\bibnamefont {Tang}},\ }\href@noop {}
  {\bibfield  {journal} {\bibinfo  {journal} {Atomic data and nuclear data
  tables}\ }\textbf {\bibinfo {volume} {56}},\ \bibinfo {pages} {159} (\bibinfo
  {year} {1994})}\BibitemShut {NoStop}%
\bibitem [{\citenamefont {Kadhane}\ \emph {et~al.}(2003)\citenamefont
  {Kadhane}, \citenamefont {Montanari},\ and\ \citenamefont
  {Tribedi}}]{kadhane2003k}%
  \BibitemOpen
  \bibfield  {author} {\bibinfo {author} {\bibfnamefont {U.}~\bibnamefont
  {Kadhane}}, \bibinfo {author} {\bibfnamefont {C.}~\bibnamefont {Montanari}},
  \ and\ \bibinfo {author} {\bibfnamefont {L.~C.}\ \bibnamefont {Tribedi}},\
  }\href@noop {} {\bibfield  {journal} {\bibinfo  {journal} {Physical Review
  A}\ }\textbf {\bibinfo {volume} {67}},\ \bibinfo {pages} {032703} (\bibinfo
  {year} {2003})}\BibitemShut {NoStop}%
\bibitem [{\citenamefont {Lapicki}(2005)}]{lapicki2005status}%
  \BibitemOpen
  \bibfield  {author} {\bibinfo {author} {\bibfnamefont {G.}~\bibnamefont
  {Lapicki}},\ }\href@noop {} {\bibfield  {journal} {\bibinfo  {journal} {X-Ray
  Spectrometry: An International Journal}\ }\textbf {\bibinfo {volume} {34}},\
  \bibinfo {pages} {269} (\bibinfo {year} {2005})}\BibitemShut {NoStop}%
\bibitem [{\citenamefont {Zhou}\ \emph {et~al.}(2013)\citenamefont {Zhou},
  \citenamefont {Zhao}, \citenamefont {Cheng}, \citenamefont {Wang},
  \citenamefont {Lei}, \citenamefont {Wang},\ and\ \citenamefont
  {Sun}}]{zhou2013k}%
  \BibitemOpen
  \bibfield  {author} {\bibinfo {author} {\bibfnamefont {X.}~\bibnamefont
  {Zhou}}, \bibinfo {author} {\bibfnamefont {Y.}~\bibnamefont {Zhao}}, \bibinfo
  {author} {\bibfnamefont {R.}~\bibnamefont {Cheng}}, \bibinfo {author}
  {\bibfnamefont {Y.}~\bibnamefont {Wang}}, \bibinfo {author} {\bibfnamefont
  {Y.}~\bibnamefont {Lei}}, \bibinfo {author} {\bibfnamefont {X.}~\bibnamefont
  {Wang}}, \ and\ \bibinfo {author} {\bibfnamefont {Y.}~\bibnamefont {Sun}},\
  }\href@noop {} {\bibfield  {journal} {\bibinfo  {journal} {Nuclear
  Instruments and Methods in Physics Research Section B: Beam Interactions with
  Materials and Atoms}\ }\textbf {\bibinfo {volume} {299}},\ \bibinfo {pages}
  {61} (\bibinfo {year} {2013})}\BibitemShut {NoStop}%
\bibitem [{\citenamefont {Msimanga}\ \emph {et~al.}(2016)\citenamefont
  {Msimanga}, \citenamefont {Pineda-Vargas},\ and\ \citenamefont
  {Madhuku}}]{msimanga2016k}%
  \BibitemOpen
  \bibfield  {author} {\bibinfo {author} {\bibfnamefont {M.}~\bibnamefont
  {Msimanga}}, \bibinfo {author} {\bibfnamefont {C.}~\bibnamefont
  {Pineda-Vargas}}, \ and\ \bibinfo {author} {\bibfnamefont {M.}~\bibnamefont
  {Madhuku}},\ }\href@noop {} {\bibfield  {journal} {\bibinfo  {journal}
  {Nuclear Instruments and Methods in Physics Research Section B: Beam
  Interactions with Materials and Atoms}\ }\textbf {\bibinfo {volume} {380}},\
  \bibinfo {pages} {90} (\bibinfo {year} {2016})}\BibitemShut {NoStop}%
\bibitem [{\citenamefont {Kumar}\ \emph {et~al.}(2017)\citenamefont {Kumar},
  \citenamefont {Singh}, \citenamefont {Oswal}, \citenamefont {Singh},
  \citenamefont {Singh}, \citenamefont {Mehta}, \citenamefont {Nandi},\ and\
  \citenamefont {Lapicki}}]{kumar2017shell}%
  \BibitemOpen
  \bibfield  {author} {\bibinfo {author} {\bibfnamefont {S.}~\bibnamefont
  {Kumar}}, \bibinfo {author} {\bibfnamefont {U.}~\bibnamefont {Singh}},
  \bibinfo {author} {\bibfnamefont {M.}~\bibnamefont {Oswal}}, \bibinfo
  {author} {\bibfnamefont {G.}~\bibnamefont {Singh}}, \bibinfo {author}
  {\bibfnamefont {N.}~\bibnamefont {Singh}}, \bibinfo {author} {\bibfnamefont
  {D.}~\bibnamefont {Mehta}}, \bibinfo {author} {\bibfnamefont
  {T.}~\bibnamefont {Nandi}}, \ and\ \bibinfo {author} {\bibfnamefont
  {G.}~\bibnamefont {Lapicki}},\ }\href@noop {} {\bibfield  {journal} {\bibinfo
   {journal} {Nuclear Instruments and Methods in Physics Research Section B:
  Beam Interactions with Materials and Atoms}\ }\textbf {\bibinfo {volume}
  {395}},\ \bibinfo {pages} {39} (\bibinfo {year} {2017})}\BibitemShut
  {NoStop}%
\bibitem [{\citenamefont {Oswal}\ \emph {et~al.}(2018)\citenamefont {Oswal},
  \citenamefont {Kumar}, \citenamefont {Singh}, \citenamefont {Singh},
  \citenamefont {Singh}, \citenamefont {Mehta}, \citenamefont {Mitnik},
  \citenamefont {Montanari},\ and\ \citenamefont {Nandi}}]{oswal2018x}%
  \BibitemOpen
  \bibfield  {author} {\bibinfo {author} {\bibfnamefont {M.}~\bibnamefont
  {Oswal}}, \bibinfo {author} {\bibfnamefont {S.}~\bibnamefont {Kumar}},
  \bibinfo {author} {\bibfnamefont {U.}~\bibnamefont {Singh}}, \bibinfo
  {author} {\bibfnamefont {G.}~\bibnamefont {Singh}}, \bibinfo {author}
  {\bibfnamefont {K.}~\bibnamefont {Singh}}, \bibinfo {author} {\bibfnamefont
  {D.}~\bibnamefont {Mehta}}, \bibinfo {author} {\bibfnamefont
  {D.}~\bibnamefont {Mitnik}}, \bibinfo {author} {\bibfnamefont {C.~C.}\
  \bibnamefont {Montanari}}, \ and\ \bibinfo {author} {\bibfnamefont
  {T.}~\bibnamefont {Nandi}},\ }\href@noop {} {\bibfield  {journal} {\bibinfo
  {journal} {Nuclear Instruments and Methods in Physics Research Section B:
  Beam Interactions with Materials and Atoms}\ }\textbf {\bibinfo {volume}
  {416}},\ \bibinfo {pages} {110} (\bibinfo {year} {2018})}\BibitemShut
  {NoStop}%
\bibitem [{\citenamefont {Hazim}\ \emph {et~al.}(2020)\citenamefont {Hazim},
  \citenamefont {Koumeir}, \citenamefont {Guertin}, \citenamefont
  {M{\'e}tivier}, \citenamefont {Naja}, \citenamefont {Servagent},\ and\
  \citenamefont {Haddad}}]{hazim2020high}%
  \BibitemOpen
  \bibfield  {author} {\bibinfo {author} {\bibfnamefont {M.}~\bibnamefont
  {Hazim}}, \bibinfo {author} {\bibfnamefont {C.}~\bibnamefont {Koumeir}},
  \bibinfo {author} {\bibfnamefont {A.}~\bibnamefont {Guertin}}, \bibinfo
  {author} {\bibfnamefont {V.}~\bibnamefont {M{\'e}tivier}}, \bibinfo {author}
  {\bibfnamefont {A.}~\bibnamefont {Naja}}, \bibinfo {author} {\bibfnamefont
  {N.}~\bibnamefont {Servagent}}, \ and\ \bibinfo {author} {\bibfnamefont
  {F.}~\bibnamefont {Haddad}},\ }\href@noop {} {\bibfield  {journal} {\bibinfo
  {journal} {Nuclear Instruments and Methods in Physics Research Section B:
  Beam Interactions with Materials and Atoms}\ }\textbf {\bibinfo {volume}
  {479}},\ \bibinfo {pages} {120} (\bibinfo {year} {2020})}\BibitemShut
  {NoStop}%
\bibitem [{\citenamefont {Oswal}\ \emph {et~al.}(2020)\citenamefont {Oswal},
  \citenamefont {Kumar}, \citenamefont {Singh}, \citenamefont {Singh},
  \citenamefont {Singh}, \citenamefont {Mehta}, \citenamefont {M{\'e}ndez},
  \citenamefont {Mitnik}, \citenamefont {Montanari}, \citenamefont {Mitra}
  \emph {et~al.}}]{oswal2020experimental}%
  \BibitemOpen
  \bibfield  {author} {\bibinfo {author} {\bibfnamefont {M.}~\bibnamefont
  {Oswal}}, \bibinfo {author} {\bibfnamefont {S.}~\bibnamefont {Kumar}},
  \bibinfo {author} {\bibfnamefont {U.}~\bibnamefont {Singh}}, \bibinfo
  {author} {\bibfnamefont {G.}~\bibnamefont {Singh}}, \bibinfo {author}
  {\bibfnamefont {K.}~\bibnamefont {Singh}}, \bibinfo {author} {\bibfnamefont
  {D.}~\bibnamefont {Mehta}}, \bibinfo {author} {\bibfnamefont
  {A.}~\bibnamefont {M{\'e}ndez}}, \bibinfo {author} {\bibfnamefont
  {D.}~\bibnamefont {Mitnik}}, \bibinfo {author} {\bibfnamefont
  {C.}~\bibnamefont {Montanari}}, \bibinfo {author} {\bibfnamefont
  {D.}~\bibnamefont {Mitra}},  \emph {et~al.},\ }\href@noop {} {\bibfield
  {journal} {\bibinfo  {journal} {Radiation Physics and Chemistry}\ ,\ \bibinfo
  {pages} {108809}} (\bibinfo {year} {2020})}\BibitemShut {NoStop}%
\bibitem [{\citenamefont {Miranda}\ \emph {et~al.}(2020)\citenamefont
  {Miranda}, \citenamefont {Serrano}, \citenamefont {Pineda}, \citenamefont
  {Mar{\'\i}n-L{\'a}mbarri}, \citenamefont {Acosta}, \citenamefont
  {Mendoza-Flores}, \citenamefont {Reynoso-Cruces},\ and\ \citenamefont
  {Ch{\'a}vez}}]{miranda2020total}%
  \BibitemOpen
  \bibfield  {author} {\bibinfo {author} {\bibfnamefont {J.}~\bibnamefont
  {Miranda}}, \bibinfo {author} {\bibfnamefont {D.}~\bibnamefont {Serrano}},
  \bibinfo {author} {\bibfnamefont {J.}~\bibnamefont {Pineda}}, \bibinfo
  {author} {\bibfnamefont {D.}~\bibnamefont {Mar{\'\i}n-L{\'a}mbarri}},
  \bibinfo {author} {\bibfnamefont {L.}~\bibnamefont {Acosta}}, \bibinfo
  {author} {\bibfnamefont {J.}~\bibnamefont {Mendoza-Flores}}, \bibinfo
  {author} {\bibfnamefont {S.}~\bibnamefont {Reynoso-Cruces}}, \ and\ \bibinfo
  {author} {\bibfnamefont {E.}~\bibnamefont {Ch{\'a}vez}},\ }\href@noop {}
  {\bibfield  {journal} {\bibinfo  {journal} {Nuclear Instruments and Methods
  in Physics Research Section B: Beam Interactions with Materials and Atoms}\
  }\textbf {\bibinfo {volume} {477}},\ \bibinfo {pages} {23} (\bibinfo {year}
  {2020})}\BibitemShut {NoStop}%
\bibitem [{\citenamefont {Lapicki}\ \emph {et~al.}(2004)\citenamefont
  {Lapicki}, \citenamefont {Murty}, \citenamefont {Raju}, \citenamefont
  {Reddy}, \citenamefont {Reddy},\ and\ \citenamefont
  {Vijayan}}]{lapicki2004effects}%
  \BibitemOpen
  \bibfield  {author} {\bibinfo {author} {\bibfnamefont {G.}~\bibnamefont
  {Lapicki}}, \bibinfo {author} {\bibfnamefont {G.~R.}\ \bibnamefont {Murty}},
  \bibinfo {author} {\bibfnamefont {G.~N.}\ \bibnamefont {Raju}}, \bibinfo
  {author} {\bibfnamefont {B.~S.}\ \bibnamefont {Reddy}}, \bibinfo {author}
  {\bibfnamefont {S.~B.}\ \bibnamefont {Reddy}}, \ and\ \bibinfo {author}
  {\bibfnamefont {V.}~\bibnamefont {Vijayan}},\ }\href@noop {} {\bibfield
  {journal} {\bibinfo  {journal} {Physical Review A}\ }\textbf {\bibinfo
  {volume} {70}},\ \bibinfo {pages} {062718} (\bibinfo {year}
  {2004})}\BibitemShut {NoStop}%
\bibitem [{\citenamefont {Dahl}\ \emph {et~al.}(1976)\citenamefont {Dahl},
  \citenamefont {Rodbro}, \citenamefont {Hermann}, \citenamefont {Fastrup},\
  and\ \citenamefont {Rudd}}]{dahl1976auger}%
  \BibitemOpen
  \bibfield  {author} {\bibinfo {author} {\bibfnamefont {P.}~\bibnamefont
  {Dahl}}, \bibinfo {author} {\bibfnamefont {M.}~\bibnamefont {Rodbro}},
  \bibinfo {author} {\bibfnamefont {G.}~\bibnamefont {Hermann}}, \bibinfo
  {author} {\bibfnamefont {B.}~\bibnamefont {Fastrup}}, \ and\ \bibinfo
  {author} {\bibfnamefont {M.}~\bibnamefont {Rudd}},\ }\href@noop {} {\bibfield
   {journal} {\bibinfo  {journal} {Journal of Physics B: Atomic and Molecular
  Physics}\ }\textbf {\bibinfo {volume} {9}},\ \bibinfo {pages} {1581}
  (\bibinfo {year} {1976})}\BibitemShut {NoStop}%
\bibitem [{\citenamefont {Pajek}\ \emph {et~al.}(2003)\citenamefont {Pajek},
  \citenamefont {Bana{\'s}}, \citenamefont {Semaniak}, \citenamefont
  {Braziewicz}, \citenamefont {Majewska}, \citenamefont {Chojnacki},
  \citenamefont {Czy{\.z}ewski}, \citenamefont {Fija{\l}}, \citenamefont
  {Jask{\'o}{\l}a}, \citenamefont {Glombik} \emph
  {et~al.}}]{pajek2003multiple}%
  \BibitemOpen
  \bibfield  {author} {\bibinfo {author} {\bibfnamefont {M.}~\bibnamefont
  {Pajek}}, \bibinfo {author} {\bibfnamefont {D.}~\bibnamefont {Bana{\'s}}},
  \bibinfo {author} {\bibfnamefont {J.}~\bibnamefont {Semaniak}}, \bibinfo
  {author} {\bibfnamefont {J.}~\bibnamefont {Braziewicz}}, \bibinfo {author}
  {\bibfnamefont {U.}~\bibnamefont {Majewska}}, \bibinfo {author}
  {\bibfnamefont {S.}~\bibnamefont {Chojnacki}}, \bibinfo {author}
  {\bibfnamefont {T.}~\bibnamefont {Czy{\.z}ewski}}, \bibinfo {author}
  {\bibfnamefont {I.}~\bibnamefont {Fija{\l}}}, \bibinfo {author}
  {\bibfnamefont {M.}~\bibnamefont {Jask{\'o}{\l}a}}, \bibinfo {author}
  {\bibfnamefont {A.}~\bibnamefont {Glombik}},  \emph {et~al.},\ }\href@noop {}
  {\bibfield  {journal} {\bibinfo  {journal} {Physical Review A}\ }\textbf
  {\bibinfo {volume} {68}},\ \bibinfo {pages} {022705} (\bibinfo {year}
  {2003})}\BibitemShut {NoStop}%
\bibitem [{\citenamefont {Sarkadi}\ and\ \citenamefont
  {Mukoyama}(1981)}]{sarkadi1981possible}%
  \BibitemOpen
  \bibfield  {author} {\bibinfo {author} {\bibfnamefont {L.}~\bibnamefont
  {Sarkadi}}\ and\ \bibinfo {author} {\bibfnamefont {T.}~\bibnamefont
  {Mukoyama}},\ }\href@noop {} {\bibfield  {journal} {\bibinfo  {journal}
  {Journal of Physics B: Atomic and Molecular Physics}\ }\textbf {\bibinfo
  {volume} {14}},\ \bibinfo {pages} {L255} (\bibinfo {year}
  {1981})}\BibitemShut {NoStop}%
\bibitem [{\citenamefont {Maidikov}\ \emph {et~al.}(1982)\citenamefont
  {Maidikov}, \citenamefont {Surovitskaya}, \citenamefont {Skobelev},\ and\
  \citenamefont {Neubert}}]{MAIDIKOV1982295}%
  \BibitemOpen
  \bibfield  {author} {\bibinfo {author} {\bibfnamefont {V.}~\bibnamefont
  {Maidikov}}, \bibinfo {author} {\bibfnamefont {N.}~\bibnamefont
  {Surovitskaya}}, \bibinfo {author} {\bibfnamefont {N.}~\bibnamefont
  {Skobelev}}, \ and\ \bibinfo {author} {\bibfnamefont {W.}~\bibnamefont
  {Neubert}},\ }\href {\doibase https://doi.org/10.1016/0029-554X(82)90836-9}
  {\bibfield  {journal} {\bibinfo  {journal} {Nuclear Instruments and Methods
  in Physics Research}\ }\textbf {\bibinfo {volume} {192}},\ \bibinfo {pages}
  {295 } (\bibinfo {year} {1982})}\BibitemShut {NoStop}%
\bibitem [{\citenamefont {Leino}\ \emph {et~al.}(1995)\citenamefont {Leino},
  \citenamefont {Äystö}, \citenamefont {Enqvist}, \citenamefont {Heikkinen},
  \citenamefont {Jokinen}, \citenamefont {Nurmia}, \citenamefont {Ostrowski},
  \citenamefont {Trzaska}, \citenamefont {Uusitalo}, \citenamefont {Eskola},
  \citenamefont {Armbruster},\ and\ \citenamefont {Ninov}}]{LEINO1995653}%
  \BibitemOpen
  \bibfield  {author} {\bibinfo {author} {\bibfnamefont {M.}~\bibnamefont
  {Leino}}, \bibinfo {author} {\bibfnamefont {J.}~\bibnamefont {Äystö}},
  \bibinfo {author} {\bibfnamefont {T.}~\bibnamefont {Enqvist}}, \bibinfo
  {author} {\bibfnamefont {P.}~\bibnamefont {Heikkinen}}, \bibinfo {author}
  {\bibfnamefont {A.}~\bibnamefont {Jokinen}}, \bibinfo {author} {\bibfnamefont
  {M.}~\bibnamefont {Nurmia}}, \bibinfo {author} {\bibfnamefont
  {A.}~\bibnamefont {Ostrowski}}, \bibinfo {author} {\bibfnamefont
  {W.}~\bibnamefont {Trzaska}}, \bibinfo {author} {\bibfnamefont
  {J.}~\bibnamefont {Uusitalo}}, \bibinfo {author} {\bibfnamefont
  {K.}~\bibnamefont {Eskola}}, \bibinfo {author} {\bibfnamefont
  {P.}~\bibnamefont {Armbruster}}, \ and\ \bibinfo {author} {\bibfnamefont
  {V.}~\bibnamefont {Ninov}},\ }\href {\doibase
  https://doi.org/10.1016/0168-583X(94)00573-7} {\bibfield  {journal} {\bibinfo
   {journal} {Nuclear Instruments and Methods in Physics Research Section B:
  Beam Interactions with Materials and Atoms}\ }\textbf {\bibinfo {volume}
  {99}},\ \bibinfo {pages} {653 } (\bibinfo {year} {1995})},\ \bibinfo {note}
  {application of Accelerators in Research and Industry '94}\BibitemShut
  {NoStop}%
\bibitem [{\citenamefont {Khuyagbaatar}\ \emph {et~al.}(2012)\citenamefont
  {Khuyagbaatar}, \citenamefont {Ackermann}, \citenamefont {Andersson},
  \citenamefont {Ballof}, \citenamefont {Brüchle}, \citenamefont {Düllmann},
  \citenamefont {Dvorak}, \citenamefont {Eberhardt}, \citenamefont {Even},
  \citenamefont {Gorshkov}, \citenamefont {Graeger}, \citenamefont
  {Heßberger}, \citenamefont {Hild}, \citenamefont {Hoischen}, \citenamefont
  {Jäger}, \citenamefont {Kindler}, \citenamefont {Kratz}, \citenamefont
  {Lahiri}, \citenamefont {Lommel}, \citenamefont {Maiti}, \citenamefont
  {Merchan}, \citenamefont {Rudolph}, \citenamefont {Schädel}, \citenamefont
  {Schaffner}, \citenamefont {Schausten}, \citenamefont {Schimpf},
  \citenamefont {Semchenkov}, \citenamefont {Serov}, \citenamefont {Türler},\
  and\ \citenamefont {Yakushev}}]{KHUYAGBAATAR201240}%
  \BibitemOpen
  \bibfield  {author} {\bibinfo {author} {\bibfnamefont {J.}~\bibnamefont
  {Khuyagbaatar}}, \bibinfo {author} {\bibfnamefont {D.}~\bibnamefont
  {Ackermann}}, \bibinfo {author} {\bibfnamefont {L.-L.}\ \bibnamefont
  {Andersson}}, \bibinfo {author} {\bibfnamefont {J.}~\bibnamefont {Ballof}},
  \bibinfo {author} {\bibfnamefont {W.}~\bibnamefont {Brüchle}}, \bibinfo
  {author} {\bibfnamefont {C.}~\bibnamefont {Düllmann}}, \bibinfo {author}
  {\bibfnamefont {J.}~\bibnamefont {Dvorak}}, \bibinfo {author} {\bibfnamefont
  {K.}~\bibnamefont {Eberhardt}}, \bibinfo {author} {\bibfnamefont
  {J.}~\bibnamefont {Even}}, \bibinfo {author} {\bibfnamefont {A.}~\bibnamefont
  {Gorshkov}}, \bibinfo {author} {\bibfnamefont {R.}~\bibnamefont {Graeger}},
  \bibinfo {author} {\bibfnamefont {F.-P.}\ \bibnamefont {Heßberger}},
  \bibinfo {author} {\bibfnamefont {D.}~\bibnamefont {Hild}}, \bibinfo {author}
  {\bibfnamefont {R.}~\bibnamefont {Hoischen}}, \bibinfo {author}
  {\bibfnamefont {E.}~\bibnamefont {Jäger}}, \bibinfo {author} {\bibfnamefont
  {B.}~\bibnamefont {Kindler}}, \bibinfo {author} {\bibfnamefont
  {J.}~\bibnamefont {Kratz}}, \bibinfo {author} {\bibfnamefont
  {S.}~\bibnamefont {Lahiri}}, \bibinfo {author} {\bibfnamefont
  {B.}~\bibnamefont {Lommel}}, \bibinfo {author} {\bibfnamefont
  {M.}~\bibnamefont {Maiti}}, \bibinfo {author} {\bibfnamefont
  {E.}~\bibnamefont {Merchan}}, \bibinfo {author} {\bibfnamefont
  {D.}~\bibnamefont {Rudolph}}, \bibinfo {author} {\bibfnamefont
  {M.}~\bibnamefont {Schädel}}, \bibinfo {author} {\bibfnamefont
  {H.}~\bibnamefont {Schaffner}}, \bibinfo {author} {\bibfnamefont
  {B.}~\bibnamefont {Schausten}}, \bibinfo {author} {\bibfnamefont
  {E.}~\bibnamefont {Schimpf}}, \bibinfo {author} {\bibfnamefont
  {A.}~\bibnamefont {Semchenkov}}, \bibinfo {author} {\bibfnamefont
  {A.}~\bibnamefont {Serov}}, \bibinfo {author} {\bibfnamefont
  {A.}~\bibnamefont {Türler}}, \ and\ \bibinfo {author} {\bibfnamefont
  {A.}~\bibnamefont {Yakushev}},\ }\href {\doibase
  https://doi.org/10.1016/j.nima.2012.06.007} {\bibfield  {journal} {\bibinfo
  {journal} {Nuclear Instruments and Methods in Physics Research Section A:
  Accelerators, Spectrometers, Detectors and Associated Equipment}\ }\textbf
  {\bibinfo {volume} {689}},\ \bibinfo {pages} {40 } (\bibinfo {year}
  {2012})}\BibitemShut {NoStop}%
\bibitem [{\citenamefont {Dickel}\ \emph {et~al.}(2015)\citenamefont {Dickel},
  \citenamefont {Plaß}, \citenamefont {{Ayet San Andres}}, \citenamefont
  {Ebert}, \citenamefont {Geissel}, \citenamefont {Haettner}, \citenamefont
  {Hornung}, \citenamefont {Miskun}, \citenamefont {Pietri}, \citenamefont
  {Purushothaman}, \citenamefont {Reiter}, \citenamefont {Rink}, \citenamefont
  {Scheidenberger}, \citenamefont {Weick}, \citenamefont {Dendooven},
  \citenamefont {Diwisch}, \citenamefont {Greiner}, \citenamefont {Heiße},
  \citenamefont {Knöbel}, \citenamefont {Lippert}, \citenamefont {Moore},
  \citenamefont {Pohjalainen}, \citenamefont {Prochazka}, \citenamefont
  {Ranjan}, \citenamefont {Takechi}, \citenamefont {Winfield},\ and\
  \citenamefont {Xu}}]{DICKEL2015137}%
  \BibitemOpen
  \bibfield  {author} {\bibinfo {author} {\bibfnamefont {T.}~\bibnamefont
  {Dickel}}, \bibinfo {author} {\bibfnamefont {W.}~\bibnamefont {Plaß}},
  \bibinfo {author} {\bibfnamefont {S.}~\bibnamefont {{Ayet San Andres}}},
  \bibinfo {author} {\bibfnamefont {J.}~\bibnamefont {Ebert}}, \bibinfo
  {author} {\bibfnamefont {H.}~\bibnamefont {Geissel}}, \bibinfo {author}
  {\bibfnamefont {E.}~\bibnamefont {Haettner}}, \bibinfo {author}
  {\bibfnamefont {C.}~\bibnamefont {Hornung}}, \bibinfo {author} {\bibfnamefont
  {I.}~\bibnamefont {Miskun}}, \bibinfo {author} {\bibfnamefont
  {S.}~\bibnamefont {Pietri}}, \bibinfo {author} {\bibfnamefont
  {S.}~\bibnamefont {Purushothaman}}, \bibinfo {author} {\bibfnamefont
  {M.}~\bibnamefont {Reiter}}, \bibinfo {author} {\bibfnamefont {A.-K.}\
  \bibnamefont {Rink}}, \bibinfo {author} {\bibfnamefont {C.}~\bibnamefont
  {Scheidenberger}}, \bibinfo {author} {\bibfnamefont {H.}~\bibnamefont
  {Weick}}, \bibinfo {author} {\bibfnamefont {P.}~\bibnamefont {Dendooven}},
  \bibinfo {author} {\bibfnamefont {M.}~\bibnamefont {Diwisch}}, \bibinfo
  {author} {\bibfnamefont {F.}~\bibnamefont {Greiner}}, \bibinfo {author}
  {\bibfnamefont {F.}~\bibnamefont {Heiße}}, \bibinfo {author} {\bibfnamefont
  {R.}~\bibnamefont {Knöbel}}, \bibinfo {author} {\bibfnamefont
  {W.}~\bibnamefont {Lippert}}, \bibinfo {author} {\bibfnamefont
  {I.}~\bibnamefont {Moore}}, \bibinfo {author} {\bibfnamefont
  {I.}~\bibnamefont {Pohjalainen}}, \bibinfo {author} {\bibfnamefont
  {A.}~\bibnamefont {Prochazka}}, \bibinfo {author} {\bibfnamefont
  {M.}~\bibnamefont {Ranjan}}, \bibinfo {author} {\bibfnamefont
  {M.}~\bibnamefont {Takechi}}, \bibinfo {author} {\bibfnamefont
  {J.}~\bibnamefont {Winfield}}, \ and\ \bibinfo {author} {\bibfnamefont
  {X.}~\bibnamefont {Xu}},\ }\href {\doibase
  https://doi.org/10.1016/j.physletb.2015.03.047} {\bibfield  {journal}
  {\bibinfo  {journal} {Physics Letters B}\ }\textbf {\bibinfo {volume}
  {744}},\ \bibinfo {pages} {137 } (\bibinfo {year} {2015})}\BibitemShut
  {NoStop}%
\bibitem [{\citenamefont {Sa’adeh}\ \emph {et~al.}(2011)\citenamefont
  {Sa’adeh}, \citenamefont {Ali},\ and\ \citenamefont
  {Arafah}}]{SAADEH20112111}%
  \BibitemOpen
  \bibfield  {author} {\bibinfo {author} {\bibfnamefont {H.}~\bibnamefont
  {Sa’adeh}}, \bibinfo {author} {\bibfnamefont {R.}~\bibnamefont {Ali}}, \
  and\ \bibinfo {author} {\bibfnamefont {D.-E.}\ \bibnamefont {Arafah}},\
  }\href {\doibase https://doi.org/10.1016/j.nimb.2011.06.020} {\bibfield
  {journal} {\bibinfo  {journal} {Nuclear Instruments and Methods in Physics
  Research Section B: Beam Interactions with Materials and Atoms}\ }\textbf
  {\bibinfo {volume} {269}},\ \bibinfo {pages} {2111 } (\bibinfo {year}
  {2011})}\BibitemShut {NoStop}%
\bibitem [{\citenamefont {Lifschitz}\ and\ \citenamefont
  {Arista}(2004)}]{lifschitz2004effective}%
  \BibitemOpen
  \bibfield  {author} {\bibinfo {author} {\bibfnamefont {A.}~\bibnamefont
  {Lifschitz}}\ and\ \bibinfo {author} {\bibfnamefont {N.}~\bibnamefont
  {Arista}},\ }\href@noop {} {\bibfield  {journal} {\bibinfo  {journal}
  {Physical Review A}\ }\textbf {\bibinfo {volume} {69}},\ \bibinfo {pages}
  {012902} (\bibinfo {year} {2004})}\BibitemShut {NoStop}%
\bibitem [{\citenamefont {Allison}(1958)}]{RevModPhys.30.1137}%
  \BibitemOpen
  \bibfield  {author} {\bibinfo {author} {\bibfnamefont {S.~K.}\ \bibnamefont
  {Allison}},\ }\href {\doibase 10.1103/RevModPhys.30.1137} {\bibfield
  {journal} {\bibinfo  {journal} {Rev. Mod. Phys.}\ }\textbf {\bibinfo {volume}
  {30}},\ \bibinfo {pages} {1137} (\bibinfo {year} {1958})}\BibitemShut
  {NoStop}%
\bibitem [{\citenamefont {Wittkower}\ and\ \citenamefont
  {Betz}(1973)}]{WITTKOWER1973113}%
  \BibitemOpen
  \bibfield  {author} {\bibinfo {author} {\bibfnamefont {A.}~\bibnamefont
  {Wittkower}}\ and\ \bibinfo {author} {\bibfnamefont {H.}~\bibnamefont
  {Betz}},\ }\href {\doibase https://doi.org/10.1016/S0092-640X(73)80001-4}
  {\bibfield  {journal} {\bibinfo  {journal} {Atomic Data and Nuclear Data
  Tables}\ }\textbf {\bibinfo {volume} {5}},\ \bibinfo {pages} {113 } (\bibinfo
  {year} {1973})}\BibitemShut {NoStop}%
\bibitem [{\citenamefont {Shima}\ \emph {et~al.}(1986)\citenamefont {Shima},
  \citenamefont {Mikumo},\ and\ \citenamefont {Tawara}}]{SHIMA1986357}%
  \BibitemOpen
  \bibfield  {author} {\bibinfo {author} {\bibfnamefont {K.}~\bibnamefont
  {Shima}}, \bibinfo {author} {\bibfnamefont {T.}~\bibnamefont {Mikumo}}, \
  and\ \bibinfo {author} {\bibfnamefont {H.}~\bibnamefont {Tawara}},\ }\href
  {\doibase https://doi.org/10.1016/0092-640X(86)90010-0} {\bibfield  {journal}
  {\bibinfo  {journal} {Atomic Data and Nuclear Data Tables}\ }\textbf
  {\bibinfo {volume} {34}},\ \bibinfo {pages} {357 } (\bibinfo {year}
  {1986})}\BibitemShut {NoStop}%
\bibitem [{\citenamefont {Shima}\ \emph {et~al.}(1992)\citenamefont {Shima},
  \citenamefont {Kuno}, \citenamefont {Yamanouchi},\ and\ \citenamefont
  {Tawara}}]{SHIMA1992173}%
  \BibitemOpen
  \bibfield  {author} {\bibinfo {author} {\bibfnamefont {K.}~\bibnamefont
  {Shima}}, \bibinfo {author} {\bibfnamefont {N.}~\bibnamefont {Kuno}},
  \bibinfo {author} {\bibfnamefont {M.}~\bibnamefont {Yamanouchi}}, \ and\
  \bibinfo {author} {\bibfnamefont {H.}~\bibnamefont {Tawara}},\ }\href
  {\doibase https://doi.org/10.1016/0092-640X(92)90001-X} {\bibfield  {journal}
  {\bibinfo  {journal} {Atomic Data and Nuclear Data Tables}\ }\textbf
  {\bibinfo {volume} {51}},\ \bibinfo {pages} {173 } (\bibinfo {year}
  {1992})}\BibitemShut {NoStop}%
\bibitem [{\citenamefont {Bohr}(1941)}]{bohr1941velocity}%
  \BibitemOpen
  \bibfield  {author} {\bibinfo {author} {\bibfnamefont {N.}~\bibnamefont
  {Bohr}},\ }\href@noop {} {\bibfield  {journal} {\bibinfo  {journal} {Physical
  Review}\ }\textbf {\bibinfo {volume} {59}},\ \bibinfo {pages} {270} (\bibinfo
  {year} {1941})}\BibitemShut {NoStop}%
\bibitem [{\citenamefont {Nikolaev}\ and\ \citenamefont
  {Dmitriev}(1968)}]{nikolaev1968equilibrium}%
  \BibitemOpen
  \bibfield  {author} {\bibinfo {author} {\bibfnamefont {V.}~\bibnamefont
  {Nikolaev}}\ and\ \bibinfo {author} {\bibfnamefont {I.}~\bibnamefont
  {Dmitriev}},\ }\href@noop {} {\bibfield  {journal} {\bibinfo  {journal}
  {Physics Letters A}\ }\textbf {\bibinfo {volume} {28}},\ \bibinfo {pages}
  {277} (\bibinfo {year} {1968})}\BibitemShut {NoStop}%
\bibitem [{\citenamefont {To}\ and\ \citenamefont
  {Drouin}(1976)}]{to1976etude}%
  \BibitemOpen
  \bibfield  {author} {\bibinfo {author} {\bibfnamefont {K.}~\bibnamefont
  {To}}\ and\ \bibinfo {author} {\bibfnamefont {R.}~\bibnamefont {Drouin}},\
  }\href@noop {} {\bibfield  {journal} {\bibinfo  {journal} {Physica Scripta}\
  }\textbf {\bibinfo {volume} {14}},\ \bibinfo {pages} {277} (\bibinfo {year}
  {1976})}\BibitemShut {NoStop}%
\bibitem [{\citenamefont {Shima}\ \emph {et~al.}(1982)\citenamefont {Shima},
  \citenamefont {Ishihara},\ and\ \citenamefont {Mikumo}}]{shima1982empirical}%
  \BibitemOpen
  \bibfield  {author} {\bibinfo {author} {\bibfnamefont {K.}~\bibnamefont
  {Shima}}, \bibinfo {author} {\bibfnamefont {T.}~\bibnamefont {Ishihara}}, \
  and\ \bibinfo {author} {\bibfnamefont {T.}~\bibnamefont {Mikumo}},\
  }\href@noop {} {\bibfield  {journal} {\bibinfo  {journal} {Nuclear
  Instruments and Methods in Physics Research}\ }\textbf {\bibinfo {volume}
  {200}},\ \bibinfo {pages} {605} (\bibinfo {year} {1982})}\BibitemShut
  {NoStop}%
\bibitem [{\citenamefont {Itoh}\ \emph {et~al.}(1999)\citenamefont {Itoh},
  \citenamefont {Tsuchida}, \citenamefont {Majima}, \citenamefont {Yogo},\ and\
  \citenamefont {Ogawa}}]{itoh1999equilibrium}%
  \BibitemOpen
  \bibfield  {author} {\bibinfo {author} {\bibfnamefont {A.}~\bibnamefont
  {Itoh}}, \bibinfo {author} {\bibfnamefont {H.}~\bibnamefont {Tsuchida}},
  \bibinfo {author} {\bibfnamefont {T.}~\bibnamefont {Majima}}, \bibinfo
  {author} {\bibfnamefont {A.}~\bibnamefont {Yogo}}, \ and\ \bibinfo {author}
  {\bibfnamefont {A.}~\bibnamefont {Ogawa}},\ }\href@noop {} {\bibfield
  {journal} {\bibinfo  {journal} {Nuclear Instruments and Methods in Physics
  Research Section B: Beam Interactions with Materials and Atoms}\ }\textbf
  {\bibinfo {volume} {159}},\ \bibinfo {pages} {22} (\bibinfo {year}
  {1999})}\BibitemShut {NoStop}%
\bibitem [{\citenamefont {Schiwietz}\ \emph {et~al.}(2004)\citenamefont
  {Schiwietz}, \citenamefont {Czerski}, \citenamefont {Roth}, \citenamefont
  {Staufenbiel},\ and\ \citenamefont {Grande}}]{schiwietz2004femtosecond}%
  \BibitemOpen
  \bibfield  {author} {\bibinfo {author} {\bibfnamefont {G.}~\bibnamefont
  {Schiwietz}}, \bibinfo {author} {\bibfnamefont {K.}~\bibnamefont {Czerski}},
  \bibinfo {author} {\bibfnamefont {M.}~\bibnamefont {Roth}}, \bibinfo {author}
  {\bibfnamefont {F.}~\bibnamefont {Staufenbiel}}, \ and\ \bibinfo {author}
  {\bibfnamefont {P.}~\bibnamefont {Grande}},\ }\href@noop {} {\bibfield
  {journal} {\bibinfo  {journal} {Nuclear Instruments and Methods in Physics
  Research Section B: Beam Interactions with Materials and Atoms}\ }\textbf
  {\bibinfo {volume} {225}},\ \bibinfo {pages} {4} (\bibinfo {year}
  {2004})}\BibitemShut {NoStop}%
\bibitem [{\citenamefont {Montanari}\ and\ \citenamefont
  {Miraglia}(2013)}]{montanari2013theory}%
  \BibitemOpen
  \bibfield  {author} {\bibinfo {author} {\bibfnamefont {C.}~\bibnamefont
  {Montanari}}\ and\ \bibinfo {author} {\bibfnamefont {J.}~\bibnamefont
  {Miraglia}},\ }\href@noop {} {\bibfield  {journal} {\bibinfo  {journal}
  {Advances in Quantum Chemistry}\ }\textbf {\bibinfo {volume} {65}},\ \bibinfo
  {pages} {165} (\bibinfo {year} {2013})}\BibitemShut {NoStop}%
\bibitem [{\citenamefont {Chatterjee}\ \emph {et~al.}(2021)\citenamefont
  {Chatterjee}, \citenamefont {Singh}, \citenamefont {Sharma}, \citenamefont
  {Oswal}, \citenamefont {Kumar}, \citenamefont {Claudia}, \citenamefont
  {Mitra},\ and\ \citenamefont {Nandi}}]{chatterjee2021investigation}%
  \BibitemOpen
  \bibfield  {author} {\bibinfo {author} {\bibfnamefont {S.}~\bibnamefont
  {Chatterjee}}, \bibinfo {author} {\bibfnamefont {S.}~\bibnamefont {Singh}},
  \bibinfo {author} {\bibfnamefont {P.}~\bibnamefont {Sharma}}, \bibinfo
  {author} {\bibfnamefont {M.}~\bibnamefont {Oswal}}, \bibinfo {author}
  {\bibfnamefont {S.}~\bibnamefont {Kumar}}, \bibinfo {author} {\bibfnamefont
  {M.}~\bibnamefont {Claudia}}, \bibinfo {author} {\bibfnamefont
  {D.}~\bibnamefont {Mitra}}, \ and\ \bibinfo {author} {\bibfnamefont
  {T.}~\bibnamefont {Nandi}},\ }\href@noop {} {\bibfield  {journal} {\bibinfo
  {journal} {arXiv preprint arXiv:2103.08299}\ } (\bibinfo {year}
  {2021})}\BibitemShut {NoStop}%
\bibitem [{\citenamefont {Krause}(1979)}]{krause1979atomic}%
  \BibitemOpen
  \bibfield  {author} {\bibinfo {author} {\bibfnamefont {M.~O.}\ \bibnamefont
  {Krause}},\ }\href@noop {} {\bibfield  {journal} {\bibinfo  {journal}
  {Journal of physical and chemical reference data}\ }\textbf {\bibinfo
  {volume} {8}},\ \bibinfo {pages} {307} (\bibinfo {year} {1979})}\BibitemShut
  {NoStop}%
\bibitem [{\citenamefont {Montanari}\ \emph {et~al.}(2011)\citenamefont
  {Montanari}, \citenamefont {Mitnik},\ and\ \citenamefont
  {Miraglia}}]{montanari2011collective}%
  \BibitemOpen
  \bibfield  {author} {\bibinfo {author} {\bibfnamefont {C.}~\bibnamefont
  {Montanari}}, \bibinfo {author} {\bibfnamefont {D.}~\bibnamefont {Mitnik}}, \
  and\ \bibinfo {author} {\bibfnamefont {J.}~\bibnamefont {Miraglia}},\
  }\href@noop {} {\bibfield  {journal} {\bibinfo  {journal} {Radiation Effects
  \& Defects in Solids}\ }\textbf {\bibinfo {volume} {166}},\ \bibinfo {pages}
  {338} (\bibinfo {year} {2011})}\BibitemShut {NoStop}%
\bibitem [{\citenamefont {Lapicki}\ and\ \citenamefont
  {Losonsky}(1977)}]{lapicki1977electron}%
  \BibitemOpen
  \bibfield  {author} {\bibinfo {author} {\bibfnamefont {G.}~\bibnamefont
  {Lapicki}}\ and\ \bibinfo {author} {\bibfnamefont {W.}~\bibnamefont
  {Losonsky}},\ }\href@noop {} {\bibfield  {journal} {\bibinfo  {journal}
  {Physical Review A}\ }\textbf {\bibinfo {volume} {15}},\ \bibinfo {pages}
  {896} (\bibinfo {year} {1977})}\BibitemShut {NoStop}%
\bibitem [{\citenamefont {May}(1964)}]{may1964formation}%
  \BibitemOpen
  \bibfield  {author} {\bibinfo {author} {\bibfnamefont {R.}~\bibnamefont
  {May}},\ }\href@noop {} {\bibfield  {journal} {\bibinfo  {journal} {PhL}\
  }\textbf {\bibinfo {volume} {11}},\ \bibinfo {pages} {33} (\bibinfo {year}
  {1964})}\BibitemShut {NoStop}%
\bibitem [{\citenamefont {Brandt}\ \emph {et~al.}(1973)\citenamefont {Brandt},
  \citenamefont {Laubert}, \citenamefont {Mourino},\ and\ \citenamefont
  {Schwarzschild}}]{brandt1973dynamic}%
  \BibitemOpen
  \bibfield  {author} {\bibinfo {author} {\bibfnamefont {W.}~\bibnamefont
  {Brandt}}, \bibinfo {author} {\bibfnamefont {R.}~\bibnamefont {Laubert}},
  \bibinfo {author} {\bibfnamefont {M.}~\bibnamefont {Mourino}}, \ and\
  \bibinfo {author} {\bibfnamefont {A.}~\bibnamefont {Schwarzschild}},\
  }\href@noop {} {\bibfield  {journal} {\bibinfo  {journal} {Physical Review
  Letters}\ }\textbf {\bibinfo {volume} {30}},\ \bibinfo {pages} {358}
  (\bibinfo {year} {1973})}\BibitemShut {NoStop}%
\bibitem [{\citenamefont {Schiwietz}\ and\ \citenamefont
  {Grande}(2001)}]{schiwietz2001improved}%
  \BibitemOpen
  \bibfield  {author} {\bibinfo {author} {\bibfnamefont {G.}~\bibnamefont
  {Schiwietz}}\ and\ \bibinfo {author} {\bibfnamefont {P.}~\bibnamefont
  {Grande}},\ }\href@noop {} {\bibfield  {journal} {\bibinfo  {journal}
  {Nuclear Instruments and Methods in Physics Research Section B: Beam
  Interactions with Materials and Atoms}\ }\textbf {\bibinfo {volume} {175}},\
  \bibinfo {pages} {125} (\bibinfo {year} {2001})}\BibitemShut {NoStop}%
\bibitem [{\citenamefont {Lamour}\ \emph {et~al.}(2015)\citenamefont {Lamour},
  \citenamefont {Fainstein}, \citenamefont {Galassi}, \citenamefont {Prigent},
  \citenamefont {Ramirez}, \citenamefont {Rivarola}, \citenamefont {Rozet},
  \citenamefont {Trassinelli},\ and\ \citenamefont {Vernhet}}]{ETACHA4}%
  \BibitemOpen
  \bibfield  {author} {\bibinfo {author} {\bibfnamefont {E.}~\bibnamefont
  {Lamour}}, \bibinfo {author} {\bibfnamefont {P.~D.}\ \bibnamefont
  {Fainstein}}, \bibinfo {author} {\bibfnamefont {M.}~\bibnamefont {Galassi}},
  \bibinfo {author} {\bibfnamefont {C.}~\bibnamefont {Prigent}}, \bibinfo
  {author} {\bibfnamefont {C.~A.}\ \bibnamefont {Ramirez}}, \bibinfo {author}
  {\bibfnamefont {R.~D.}\ \bibnamefont {Rivarola}}, \bibinfo {author}
  {\bibfnamefont {J.-P.}\ \bibnamefont {Rozet}}, \bibinfo {author}
  {\bibfnamefont {M.}~\bibnamefont {Trassinelli}}, \ and\ \bibinfo {author}
  {\bibfnamefont {D.}~\bibnamefont {Vernhet}},\ }\href {\doibase
  10.1103/PhysRevA.92.042703} {\bibfield  {journal} {\bibinfo  {journal} {Phys.
  Rev. A}\ }\textbf {\bibinfo {volume} {92}},\ \bibinfo {pages} {042703}
  (\bibinfo {year} {2015})}\BibitemShut {NoStop}%
\bibitem [{\citenamefont {Meyerhof}\ \emph {et~al.}(1985)\citenamefont
  {Meyerhof}, \citenamefont {Anholt}, \citenamefont {Eichler}, \citenamefont
  {Gould}, \citenamefont {Munger}, \citenamefont {Alonso}, \citenamefont
  {Thieberger},\ and\ \citenamefont {Wegner}}]{Meyerhof}%
  \BibitemOpen
  \bibfield  {author} {\bibinfo {author} {\bibfnamefont {W.~E.}\ \bibnamefont
  {Meyerhof}}, \bibinfo {author} {\bibfnamefont {R.}~\bibnamefont {Anholt}},
  \bibinfo {author} {\bibfnamefont {J.}~\bibnamefont {Eichler}}, \bibinfo
  {author} {\bibfnamefont {H.}~\bibnamefont {Gould}}, \bibinfo {author}
  {\bibfnamefont {C.}~\bibnamefont {Munger}}, \bibinfo {author} {\bibfnamefont
  {J.}~\bibnamefont {Alonso}}, \bibinfo {author} {\bibfnamefont
  {P.}~\bibnamefont {Thieberger}}, \ and\ \bibinfo {author} {\bibfnamefont
  {H.~E.}\ \bibnamefont {Wegner}},\ }\href {\doibase 10.1103/PhysRevA.32.3291}
  {\bibfield  {journal} {\bibinfo  {journal} {Phys. Rev. A}\ }\textbf {\bibinfo
  {volume} {32}},\ \bibinfo {pages} {3291} (\bibinfo {year}
  {1985})}\BibitemShut {NoStop}%
\bibitem [{\citenamefont {Bethe}\ and\ \citenamefont
  {Salpeter}(1957)}]{bethe1957quantum}%
  \BibitemOpen
  \bibfield  {author} {\bibinfo {author} {\bibfnamefont {H.~A.}\ \bibnamefont
  {Bethe}}\ and\ \bibinfo {author} {\bibfnamefont {E.~E.}\ \bibnamefont
  {Salpeter}},\ }\href@noop {} {\bibfield  {journal} {\bibinfo  {journal}
  {Encyclopedia of Physics}\ }\textbf {\bibinfo {volume} {35}} (\bibinfo {year}
  {1957})}\BibitemShut {NoStop}%
\bibitem [{\citenamefont {Fainstein}\ \emph {et~al.}(1987)\citenamefont
  {Fainstein}, \citenamefont {Ponce},\ and\ \citenamefont
  {Rivarola}}]{fainstein1987z}%
  \BibitemOpen
  \bibfield  {author} {\bibinfo {author} {\bibfnamefont {P.~D.}\ \bibnamefont
  {Fainstein}}, \bibinfo {author} {\bibfnamefont {V.~H.}\ \bibnamefont
  {Ponce}}, \ and\ \bibinfo {author} {\bibfnamefont {R.~D.}\ \bibnamefont
  {Rivarola}},\ }\href@noop {} {\bibfield  {journal} {\bibinfo  {journal}
  {Physical Review A}\ }\textbf {\bibinfo {volume} {36}},\ \bibinfo {pages}
  {3639} (\bibinfo {year} {1987})}\BibitemShut {NoStop}%
\bibitem [{\citenamefont {Fainstein}\ \emph {et~al.}(1991)\citenamefont
  {Fainstein}, \citenamefont {Ponce},\ and\ \citenamefont
  {Rivarola}}]{fainstein1991two}%
  \BibitemOpen
  \bibfield  {author} {\bibinfo {author} {\bibfnamefont {P.}~\bibnamefont
  {Fainstein}}, \bibinfo {author} {\bibfnamefont {V.~H.}\ \bibnamefont
  {Ponce}}, \ and\ \bibinfo {author} {\bibfnamefont {R.~D.}\ \bibnamefont
  {Rivarola}},\ }\href@noop {} {\bibfield  {journal} {\bibinfo  {journal}
  {Journal of Physics B: Atomic, Molecular and Optical Physics}\ }\textbf
  {\bibinfo {volume} {24}},\ \bibinfo {pages} {3091} (\bibinfo {year}
  {1991})}\BibitemShut {NoStop}%
\bibitem [{\citenamefont {Olivera}\ \emph {et~al.}(1993)\citenamefont
  {Olivera}, \citenamefont {Ram{\'\i}rez},\ and\ \citenamefont
  {Rivarola}}]{olivera1993electronic}%
  \BibitemOpen
  \bibfield  {author} {\bibinfo {author} {\bibfnamefont {G.~H.}\ \bibnamefont
  {Olivera}}, \bibinfo {author} {\bibfnamefont {C.~A.}\ \bibnamefont
  {Ram{\'\i}rez}}, \ and\ \bibinfo {author} {\bibfnamefont {R.~D.}\
  \bibnamefont {Rivarola}},\ }\href@noop {} {\bibfield  {journal} {\bibinfo
  {journal} {Physical Review A}\ }\textbf {\bibinfo {volume} {47}},\ \bibinfo
  {pages} {1000} (\bibinfo {year} {1993})}\BibitemShut {NoStop}%
\bibitem [{\citenamefont {Ram{\'\i}rez}\ and\ \citenamefont
  {Rivarola}(1995)}]{ramirez1995distortion}%
  \BibitemOpen
  \bibfield  {author} {\bibinfo {author} {\bibfnamefont {C.~A.}\ \bibnamefont
  {Ram{\'\i}rez}}\ and\ \bibinfo {author} {\bibfnamefont {R.~D.}\ \bibnamefont
  {Rivarola}},\ }\href@noop {} {\bibfield  {journal} {\bibinfo  {journal}
  {Physical Review A}\ }\textbf {\bibinfo {volume} {52}},\ \bibinfo {pages}
  {4972} (\bibinfo {year} {1995})}\BibitemShut {NoStop}%
\bibitem [{\citenamefont {Tolk}\ \emph {et~al.}(1981)\citenamefont {Tolk},
  \citenamefont {Feldman}, \citenamefont {Kraus}, \citenamefont {Tully},
  \citenamefont {Hass}, \citenamefont {Niv},\ and\ \citenamefont
  {Temmer}}]{tolk1981role}%
  \BibitemOpen
  \bibfield  {author} {\bibinfo {author} {\bibfnamefont {N.}~\bibnamefont
  {Tolk}}, \bibinfo {author} {\bibfnamefont {L.~C.}\ \bibnamefont {Feldman}},
  \bibinfo {author} {\bibfnamefont {J.}~\bibnamefont {Kraus}}, \bibinfo
  {author} {\bibfnamefont {J.}~\bibnamefont {Tully}}, \bibinfo {author}
  {\bibfnamefont {M.}~\bibnamefont {Hass}}, \bibinfo {author} {\bibfnamefont
  {Y.}~\bibnamefont {Niv}}, \ and\ \bibinfo {author} {\bibfnamefont
  {G.}~\bibnamefont {Temmer}},\ }\href@noop {} {\bibfield  {journal} {\bibinfo
  {journal} {Physical Review Letters}\ }\textbf {\bibinfo {volume} {47}},\
  \bibinfo {pages} {487} (\bibinfo {year} {1981})}\BibitemShut {NoStop}%
\bibitem [{\citenamefont {Gall}(2016)}]{gall2016electron}%
  \BibitemOpen
  \bibfield  {author} {\bibinfo {author} {\bibfnamefont {D.}~\bibnamefont
  {Gall}},\ }\href@noop {} {\bibfield  {journal} {\bibinfo  {journal} {Journal
  of Applied Physics}\ }\textbf {\bibinfo {volume} {119}},\ \bibinfo {pages}
  {085101} (\bibinfo {year} {2016})}\BibitemShut {NoStop}%
\bibitem [{\citenamefont {Isaacson}(1975)}]{isaacson1975compilation}%
  \BibitemOpen
  \bibfield  {author} {\bibinfo {author} {\bibfnamefont {D.}~\bibnamefont
  {Isaacson}},\ }\href@noop {} {\bibfield  {journal} {\bibinfo  {journal} {New
  York University, Document}\ } (\bibinfo {year} {1975})}\BibitemShut {NoStop}%
\bibitem [{\citenamefont {Novikov}\ and\ \citenamefont
  {Teplova}(2014)}]{novikov2014method}%
  \BibitemOpen
  \bibfield  {author} {\bibinfo {author} {\bibfnamefont {N.}~\bibnamefont
  {Novikov}}\ and\ \bibinfo {author} {\bibfnamefont {Y.~A.}\ \bibnamefont
  {Teplova}},\ }\href@noop {} {\bibfield  {journal} {\bibinfo  {journal}
  {Physics Letters A}\ }\textbf {\bibinfo {volume} {378}},\ \bibinfo {pages}
  {1286} (\bibinfo {year} {2014})}\BibitemShut {NoStop}%
\bibitem [{\citenamefont {Fija{\l}-Kirejczyk}\ \emph
  {et~al.}(2008)\citenamefont {Fija{\l}-Kirejczyk}, \citenamefont
  {Jask{\'o}{\l}a}, \citenamefont {Czarnacki}, \citenamefont {Korman},
  \citenamefont {Bana{\'s}}, \citenamefont {Braziewicz}, \citenamefont
  {Majewska}, \citenamefont {Semaniak}, \citenamefont {Pajek}, \citenamefont
  {Kretschmer} \emph {et~al.}}]{fijal2008coupling}%
  \BibitemOpen
  \bibfield  {author} {\bibinfo {author} {\bibfnamefont {I.}~\bibnamefont
  {Fija{\l}-Kirejczyk}}, \bibinfo {author} {\bibfnamefont {M.}~\bibnamefont
  {Jask{\'o}{\l}a}}, \bibinfo {author} {\bibfnamefont {W.}~\bibnamefont
  {Czarnacki}}, \bibinfo {author} {\bibfnamefont {A.}~\bibnamefont {Korman}},
  \bibinfo {author} {\bibfnamefont {D.}~\bibnamefont {Bana{\'s}}}, \bibinfo
  {author} {\bibfnamefont {J.}~\bibnamefont {Braziewicz}}, \bibinfo {author}
  {\bibfnamefont {U.}~\bibnamefont {Majewska}}, \bibinfo {author}
  {\bibfnamefont {J.}~\bibnamefont {Semaniak}}, \bibinfo {author}
  {\bibfnamefont {M.}~\bibnamefont {Pajek}}, \bibinfo {author} {\bibfnamefont
  {W.}~\bibnamefont {Kretschmer}},  \emph {et~al.},\ }\href@noop {} {\bibfield
  {journal} {\bibinfo  {journal} {Physical Review A}\ }\textbf {\bibinfo
  {volume} {77}},\ \bibinfo {pages} {032706} (\bibinfo {year}
  {2008})}\BibitemShut {NoStop}%
\bibitem [{\citenamefont {Grigorenko}\ \emph {et~al.}(2019)\citenamefont
  {Grigorenko}, \citenamefont {Sharkov}, \citenamefont {Fomichev},
  \citenamefont {Barabanov}, \citenamefont {Barth}, \citenamefont {Bezbakh},
  \citenamefont {Bogomolov}, \citenamefont {Golovkov}, \citenamefont
  {Gorshkov}, \citenamefont {Dmitriev} \emph
  {et~al.}}]{grigorenko2019scientific}%
  \BibitemOpen
  \bibfield  {author} {\bibinfo {author} {\bibfnamefont {L.~V.}\ \bibnamefont
  {Grigorenko}}, \bibinfo {author} {\bibfnamefont {B.~Y.}\ \bibnamefont
  {Sharkov}}, \bibinfo {author} {\bibfnamefont {A.~S.}\ \bibnamefont
  {Fomichev}}, \bibinfo {author} {\bibfnamefont {A.~L.}\ \bibnamefont
  {Barabanov}}, \bibinfo {author} {\bibfnamefont {W.}~\bibnamefont {Barth}},
  \bibinfo {author} {\bibfnamefont {A.}~\bibnamefont {Bezbakh}}, \bibinfo
  {author} {\bibfnamefont {S.}~\bibnamefont {Bogomolov}}, \bibinfo {author}
  {\bibfnamefont {M.~S.}\ \bibnamefont {Golovkov}}, \bibinfo {author}
  {\bibfnamefont {A.}~\bibnamefont {Gorshkov}}, \bibinfo {author}
  {\bibfnamefont {S.~N.}\ \bibnamefont {Dmitriev}},  \emph {et~al.},\
  }\href@noop {} {\bibfield  {journal} {\bibinfo  {journal} {Physics-Uspekhi}\
  }\textbf {\bibinfo {volume} {62}},\ \bibinfo {pages} {675} (\bibinfo {year}
  {2019})}\BibitemShut {NoStop}%
\bibitem [{\citenamefont {Adonin}\ and\ \citenamefont
  {Hollinger}(2014)}]{adonin2014beam}%
  \BibitemOpen
  \bibfield  {author} {\bibinfo {author} {\bibfnamefont {A.}~\bibnamefont
  {Adonin}}\ and\ \bibinfo {author} {\bibfnamefont {R.}~\bibnamefont
  {Hollinger}},\ }\href@noop {} {\bibfield  {journal} {\bibinfo  {journal}
  {Review of Scientific Instruments}\ }\textbf {\bibinfo {volume} {85}},\
  \bibinfo {pages} {02A727} (\bibinfo {year} {2014})}\BibitemShut {NoStop}%
\end{thebibliography}%
\bibliographystyle{apsrev4-1}
\end{document}